\documentclass{article}
\usepackage{amsmath,amssymb,mathrsfs}
\usepackage{amsthm}
\usepackage[colorlinks]{hyperref}
\usepackage[capitalise]{cleveref}
\usepackage{url}
\usepackage{graphicx,color,epsfig}
\usepackage{epstopdf}
\usepackage{amsfonts}
\usepackage{caption}
\usepackage{enumerate}
\usepackage{enumitem}
\usepackage{float}
\usepackage{booktabs}
\usepackage{graphicx}
\usepackage{caption}
\usepackage{subcaption}
\usepackage{array}
\usepackage{booktabs}
\usepackage{fancyhdr}
\usepackage{eurosym}
\usepackage{cite}
\usepackage{upgreek}
\usepackage{enumitem}
\newlist{steps}{enumerate}{1}
\setlist[steps, 1]{label = Step \arabic*:}
\usepackage{multicol}
\hypersetup{pdftitle={},
    pdfauthor={Lanre Akinyemi},     
    pdfcreator={pdftex},   
    pdfproducer={Texmaker}, 
    pdfkeywords={}, 
    colorlinks=true,       
    linkcolor=blue,          
    citecolor=blue,        
    filecolor=magenta,      
    urlcolor=blue           
}
\DeclareMathOperator{\sech}{sech}

\Crefname{section}{Sect.}{Sects.}
\Crefname{figure}{Fig.}{Figs.}
\Crefname{table}{Tab.}{Tabs.}
\makeatletter
\let\oldtheequation\theequation
\renewcommand\tagform@[1]{\maketag@@@{\ignorespaces#1\unskip\@@italiccorr}}
\renewcommand\theequation{(\oldtheequation)}
\makeatother

\date{}

\begin{document}
\title{Modulation instability gain and localized waves by modified Frenkel-Kontorova model of higher order nonlinearity}
\author{Alphonse Houwe$^{1,\ast}\,$ Souleymanou Abbagari$^{2,\ast},\,$ Lanre Akinyemi$^{3,\ast}\,$ Serge Yamigno Doka$^{4},\,$\\ Kofan\'{e} Timol\'{e}on Cr\'{e}pin$^{5}$\\
\\
$^{1}$\textit{Department of Marine Engineering, Limb\'{e} Nautical Arts and Fisheries Institute,}\\
\textit{P.O Box 485, Limb\'{e}, Cameroon}\\
$^{2}$\textit{Department of Basic Science, National Advanced School of Mines and Petroleum Industries,}\\
\textit{The University of Maroua, P.O Box 08, ka\'{e}l\'{e}, Cameroon}\\
$^{3}$\textit{Department of Mathematics, Lafayette College, Easton, Pennsylvania, USA}\\
$^{4}$\textit{Department of Physics, Faculty of Science, The University of Ngaound\'{e}r\'{e},}\\
\textit{P.O. Box 454, Ngaound\'{e}r\'{e}, Cameroon.}\\
$^{5}$\textit{Department of Physics and Astronomy, Botswana International University of Science and Technology,}\\
 \textit{Private Mail Bag 16, Palapye, Botswana}.\\
 \\
\textit{e-mail: ahouw220@yahoo.fr,\,abbagaris@yahoo.fr,\,akinyeml@lafayette.edu,}\\
\textit{numami@gmail.com,\,tckofane@yahoo.com}\\
\\
\textit{Corresponding authors:$\,\,\,^*$}}
\maketitle
\begin{abstract}
\noindent
In this paper, modulation instability and nonlinear supratransmission are investigated in a one-dimensional chain of atoms using cubic-quartic nonlinearity coefficients. As a result, we establish the discrete nonlinear evolution equation by using the multi-scale scheme. To calculate the modulation instability gain, we use the linearizing scheme. Particular attention is given to the impact of the higher nonlinear term on the modulation instability. Following that, full numerical integration was performed to identify modulated wave patterns, as well as the appearance of a rogue wave. Through the nonlinear supratransmission phenomenon, one end of the discrete model is driven into the forbidden bandgap. As a result, for driving amplitudes above the supratransmission threshold, the solitonic bright soliton and modulated wave patterns are satisfied. An important behavior is observed in the transient range of time of propagation when the bright solitonic wave turns into a chaotic solitonic wave. These results corroborate our analytical investigations on the modulation instability and show that the one-dimensional chain of atoms is a fruitful medium to generate long-lived modulated waves.\\

\noindent
\textbf{Keywords:} Modified Frenkel-Kontorova model; Modulation instability; Modulated waves patterns; Rogue waves
\end{abstract}
\section{Introduction}
\noindent
In recent years, investigation of the localized waves in nonlinear systems has grown. A wide class of nonlinear evolution equations have been employed in different fields such as optical fibers, Bose-Einstein condensates, optomechanical lattices, molecular chains, fluid mechanics, and ferromagnetic structures \cite{H1,H2,H3,H3a,H3b,H4,H5,H5a,H5b,H6,H7,H8,H9,H10,H11,H12,H13,H14,H15,H16,H17,H18,H19,H20,H21}. More often, the phenomenon that exhibits the behavior of the excited localized modes is the modulation instability (MI). MI is characterized by a rapidly growing plane wave amplitude subject to small perturbations where nonlinear and dispersion terms interplay \cite{H2,H4,H5,H9,H11,H13,H19}. Usually, MI is manifested by the unstable or stable zones generated by the perturbed wave vector or nonlinear strength, which leads to the formation of modulated wave (MW) patterns. The three types of localized excitations that can be obtained are bright solitons, rogue waves (RWs), and breathers. For example, Conrad and co-workers have recently pointed out the propagation of the RWs of  A and B types in a nonlocal medium where a nonlinear saturation parameter is used \cite{H3a}. The authors have shown that when the MI is developed,  the MW patterns emerge. In \cite{H3b}, the authors have exhibited the localized modes in a nonlinear oscillator lattice through the MI. It is obvious that the MI is an appropriate mechanism for the generation of localized waves. If most of the models used before were in the continuum limit, it is evident today that the discrete MI of the continuous wave (CW) in the discrete nonlinear evolution equation (DNLEE) has gained a lot of interest. Houwe et al. have used the DNLEE, which describes the wave propagating in the ferromagnetic chains, to develop the MI under the effects of the nearest-neighbor coupling. The discrete MI has been the subject of theoretical and experimental research. The work of Mounouna et al. \cite{H5} is a well-known example of the MI growth rate, where the effects of the nonlinear cubic-quartic coupling of the modified Frenkel-Kontorova model were shown on the gain profile and the unstable zones that emerged during the long time of plane wave propagation.\\

\noindent
If the MI is an important process for developing localized waves, nonlinear supratransmission also remains a powerful tool where energy can flow in the forbidden bandgap (FG). This phenomenon has been developed by F. Geniet and co-workers by using the Klein Gordon equation. They have shown that when the amplitude of the driving is considered above the threshold for supratransmission, energy can flow in the FG \cite{H7}. Khomeriki et al. used the same procedure to derive the static breather solution, which synchronizes and adjusts to the Fermi-Pasta-Ulam model \cite{H8}. Beside this, several other studies have been developed to show that nonlinear strength can also favor the formation of the MWs patterns in the FG \cite{H9,H18,H21}.\\

\noindent
Very recently, the supratransmission has been exhibited by driving one end of the chain where the on-site potential of a cubic form has been used \cite{H6}. It has been called the quartic nonlinear interaction potential. Nonlinear supratransmission has been used in other applications, such as three-wave sharp interaction and low-light pulses when a two-level medium produces solitary waves \cite{H16}. In the present study, we point out the MWs, RWs, and diverse other localized waves under the effects of the cubic and quartic nonlinear interaction potentials. Thereafter, we subject one end of the chain to an external periodic force to demonstrate the supratransmission phenomenon. It emerges that, above the threshold, supratransmission of a localized bright soliton is fulfilled. We equally observed that when the driven amplitude (DA) is strong enough, the transient regime manifests itself by the escape of the bright soliton to chaos-like motion.\\

\noindent
The rest of the paper is sketched as follows: In \Cref{sec2}, we present the proposed model and thereafter use the standard multi-scale scheme to derive the DNLEE. \Cref{sec3} gives the linear stability of the MI. An expression of the MI growth rate is deduced from the dispersion relation and used to show the unstable and stable zones. In particular, we focused on the dispersion coefficient and the impact of the cubic and quadratic nonlinear interactions. \Cref{sec4} uses the full numerical simulation to corroborate analytical predictions, MW patterns, and RWs. On the other hand, one end of the chain is driven by an external periodic force. In FG, the formation of excited localized modes and bright soliton is observed in equal measure. \Cref{sec5} is assigned to concluding the work.
\section{Analytical model}\label{sec2}
Motivated by the work of Mounouna et al. \cite{H5}, we consider in this work a chain of coupled atoms subjugated to a deformable external periodic potential where the total Hamiltonian is written as:
\begin{equation}\label{eq:01}
\begin{gathered}
\mathbf{H}=\Gamma\sum_n \Bigg[\frac{1}{2}\left(\frac{d\theta_n}{dt}\right)^2+\bigg(\frac{1}{2}G_2(\theta_n-\theta_{n-1})^2+\frac{1}{3}G_3(\theta_n-\theta_{n-1})^3+\frac{1}{4}G_4(\theta_n-\theta_{n-1})^4\bigg)+\omega^2_0\frac{\tan^2\left(\frac{\theta_n}{2}\right)}{\left(\sigma^2+\tan^2\left(\frac{\theta_n}{2}\right)\right)}\Bigg],
\end{gathered}
\end{equation}
\noindent
where $\Gamma$ denotes the energy scale, $\theta_n$ the dimensionless movements of particles and $\omega_g$ the angular frequency. The potential interaction coefficients are $G_j (j=2,3,4)$ and the equation of the motion reads \cite{H5}:
\begin{equation}\label{eq:02}
\begin{gathered}
\frac{d^2\theta_n}{dt^2}=G_2(\theta_{n+1}-2\theta_n+\theta_{n-1})+G_3\left[(\theta_{n+1}-\theta_n)^2-(\theta_{n}-\theta_{n-1})^2\right]+G_4\left[(\theta_{n+1}-\theta_n)^3-(\theta_{n}-\theta_{n-1})^3\right]\\
-\omega^2_g(\theta_n+\alpha\theta^2_n+\beta\theta^3_n).
\end{gathered}
\end{equation}
\noindent
The frequency parameter is $\omega_g$, the nonlinearities in the potential's shape are $\alpha$ and $\beta$. \autoref{eq:02} is the discrete equation that describes the movement of the chains of particles in the presence of the nonlinear coupling terms, and it takes its origin from the modified Frenkel-Kontorova model. In \cite{H5a}, the authors have considered the model with $G_3=0$ and $G_4=0$ to exhibit the effects of the nonlinearity coefficients and the substrate's deformability on MI. The model was recently used to investigate the interaction between cubic-quartic nonlinearities and the substrate's deformable parameter on MI growth rates \cite{H5}. The authors have shown that the influence of the quartic nonlinearity has extended the MI bands and that the amplitude of the plane wave has risen exponentially. It is important to underline that \autoref{eq:02} can take the form of  the Klein Gordon equation when cubic-quartic interactions and coupling are omitted. In what follows, we aim to highlight the effect of the nonlinearity coefficients on the MI growth rates. Thereafter, we establish the threshold amplitude (TA) expression, from which we will consider the DA to drive one end of the model in the forbidden frequency band gap. A fascinating matter that has been studied in several nonlinear systems is driven at one end of the model. But, to our knowledge, this subject was not formulated in \cite{H5}. To do so, we consider the standard multi-scale analytical paths as follows:
\begin{equation}\label{eq:03}
\begin{gathered}
\theta_n(t)=\epsilon\left(\psi_n(\epsilon^2,t)e^{i\phi_n}+cc\right)+\epsilon^2\Theta_n(\epsilon^2,t)+\epsilon\left(\Gamma_n(\epsilon^2,t)e^{2i\phi_n}+cc\right).
\end{gathered}
\end{equation}
\noindent
Here, \autoref{eq:03} is the slowly varying time and space of MW solution, which propagates at a carrier frequency $\omega$ and wave vector $q$, $\epsilon$ counts for the small parameter and the phase is $\phi_n=kn-\omega{t}.$ While $\psi_n$ and $\Gamma_n$ are the complex functions with cc denotes their complex conjugate, $\Theta_n$ is a real function. Assuming that $G_2\sim\epsilon^2,\,$ $G_3\sim1,\,$ $G_4\sim1,\,$ $\alpha\sim1$ and $\beta\sim1$ \cite{H5,H6}. Inserting \autoref{eq:03} into  \autoref{eq:02} and gathering
terms in order $\epsilon,\,$  $\epsilon^2$ and  $\epsilon^3$ together with $e^{2i\phi_n}$, we get the DNLEE:
 \begin{equation}\label{eq:04}
\Theta_n=2\alpha|\psi_n|^2,\,\,\,\,\Gamma_n=-\frac{\alpha\omega^2_g\psi^2_n}{4\omega^2-\omega^2_g},
\end{equation}
\noindent
and
\begin{equation}\label{eq:05}
\begin{gathered}
-2i\omega\dot{\psi}_n+(\omega^2_g-\omega^2)\psi_n-G_2(\psi_{n+1}e^{ik}-2\psi_n+\psi_{n-1}e^{-ik})+3G_3\left(\psi^\ast_{n-1}e^{ik}-\psi_n^\ast\right)\left(\psi_{n-1}e^{-ik}-\psi_n\right)^2\\
-3G_4\left[\left(\psi^\ast_{n-1}e^{ik}-\psi_n^\ast\right)\left(\psi_{n-1}e^{-ik}-\psi_n\right)^2+\left(\psi^\ast_{n+1}e^{-ik}-\psi_n^\ast\right)\left(\psi_{n+1}e^{ik}-\psi_n\right)^2\right]\\
+\omega^2_g(3\beta-4\alpha^2+\frac{2\omega^2_g\alpha^2}{4\omega^2-\omega^2_g})|\psi_n|^2\psi_n=0.
\end{gathered}
\end{equation}
Thus, \autoref{eq:05} is the DNLEE with cubic-quartic interaction potential. As we have mentioned above, the upcoming section will discuss the MI growth rates.
\section{Modulation instability}\label{sec3}
\begin{figure}[H]
\begin{center}
\includegraphics[width=17cm,height=9cm]{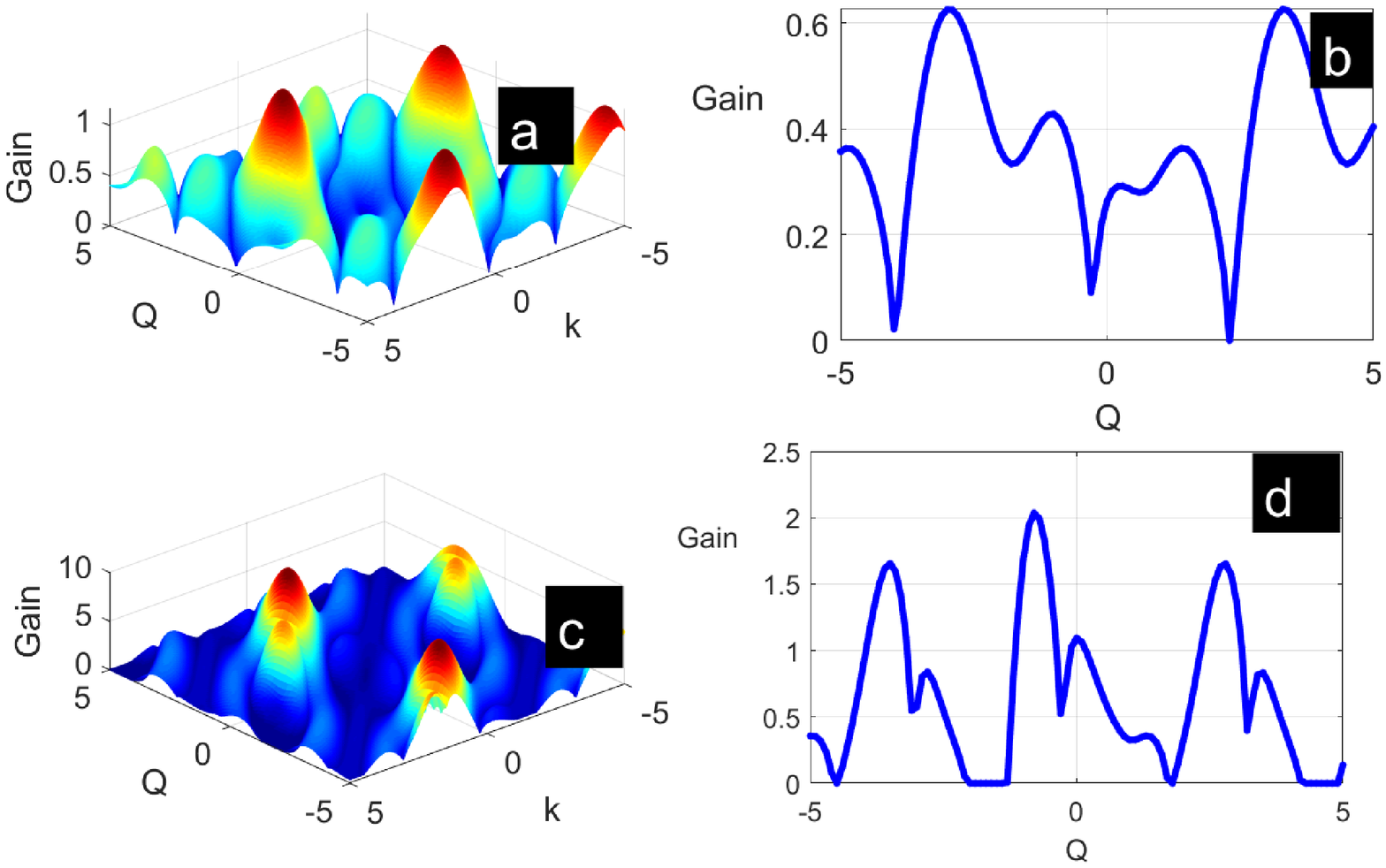}
\includegraphics[width=17cm,height=9cm]{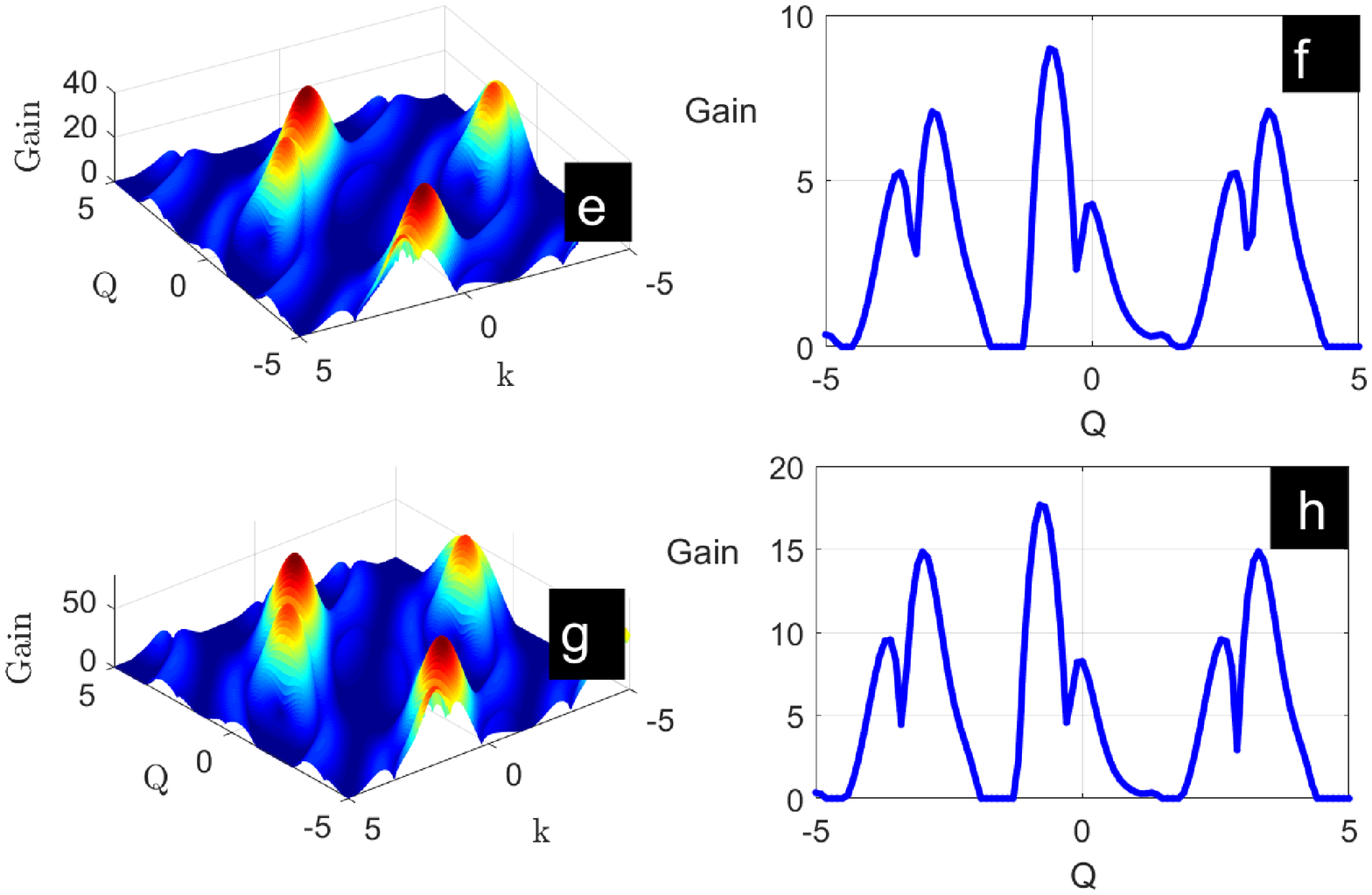}
\caption{Illustration of the MI growth rate with the variation of the quartic interaction potential. (a-b) $G_4=-0.01,\,$ (c-d) $G_4=-0.1,\,$ (e-f) $G_4=-0.5,\,$ and (g-h) $G_4=-1.$ The parameters used are $G_2=0.01,\, G_3=0,\, \alpha=1.5,\, \beta=-\frac{1}{6},\, \omega_g=0.24,\, \omega=0.42,$ and $F_0=1.$}\label{fig1}
\end{center}
\end{figure}

\begin{figure}[H]
\begin{center}
\includegraphics[width=17cm,height=9cm]{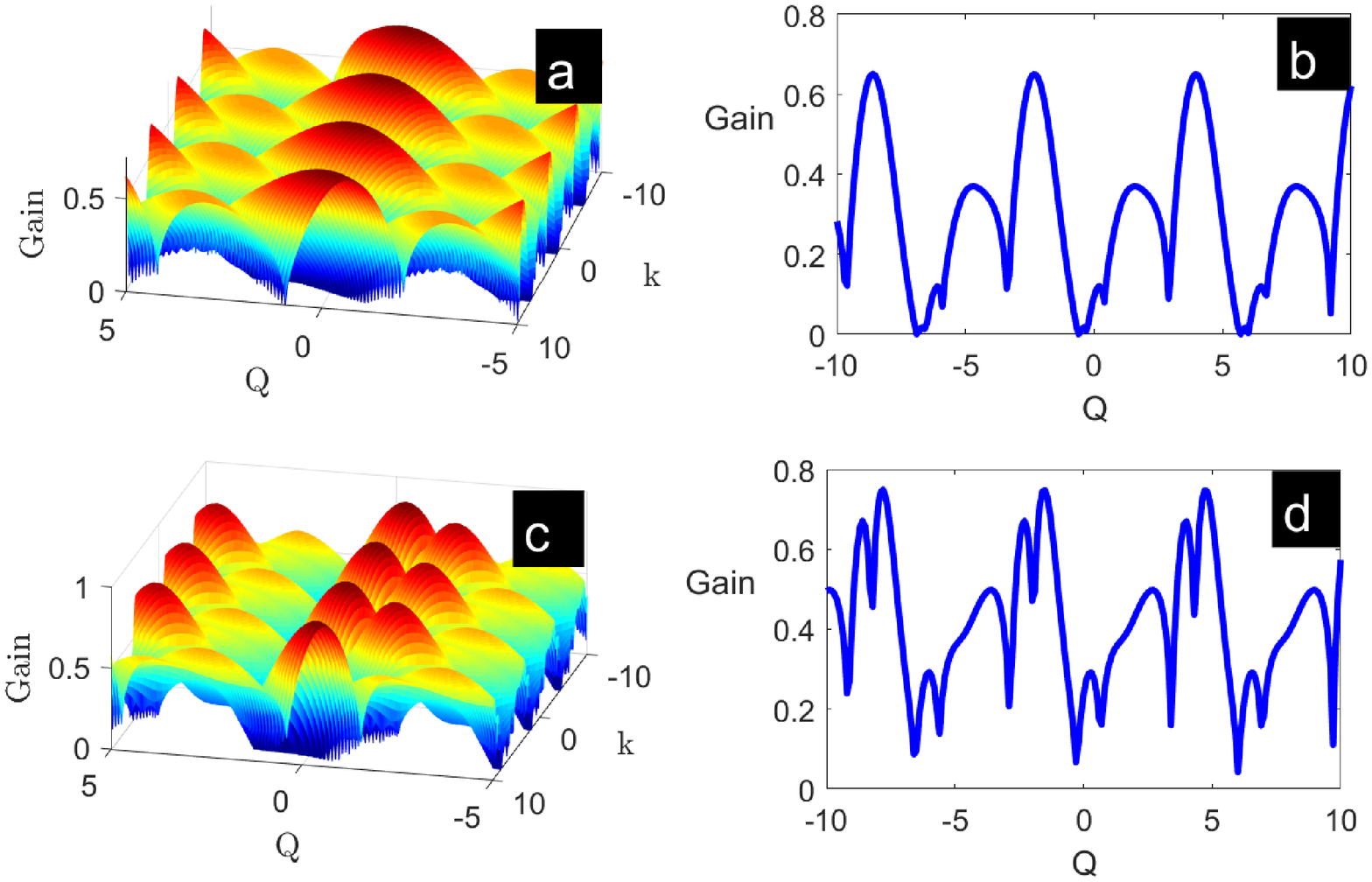}
\includegraphics[width=17cm,height=9cm]{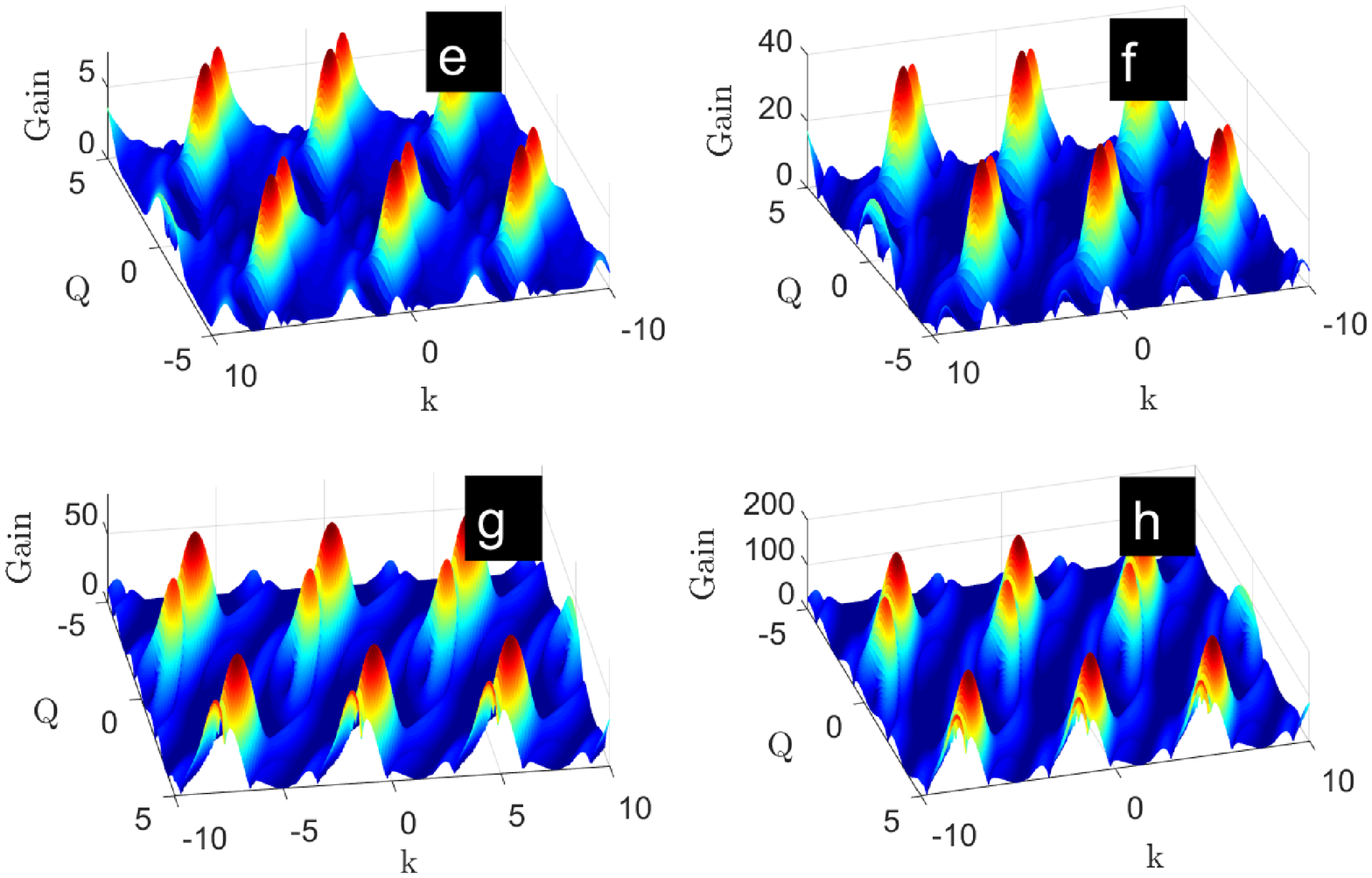}
\caption{Variation of the MI growth rate under the positive value of the quartic nonlinearity strength. (a-b) $G_4=0.01,\,$ (c-d) $G_4=0.1,\,$ (e) $G_4=0.5,\,$ (f) $G_4=1,\,$  (g) $G_4=1.5,\,$ and (h) $G_4=2.5.$ The parameters used are $G_2=0.01,\, G_3=0,\, \alpha=1.5,\, \beta=-\frac{1}{6},\, \omega_g=0.24,\, \omega=0.42,$ and $F_0=1.$ }\label{fig2}
\end{center}
\end{figure}

MI is the phenomenon where nonlinearity and dispersion interplay. Some works has been reported in \cite{H2,H4,H9,H11,H15}. During the investigation of the MI, the DNLEEs are the models that are most widely involved. For example, in \cite{H2}, the authors used the well-known discrete nonlinear Schr\"{o}dinger equation with cubic-quintic nonlinearity to analyze the MI growth rates under the effects of the nonlinear term. Tanemura et al. investigate the modulational unstable or stable modes in loss-dispersion optical fibers. More recently, the effects of the nearest-neighbor coupling have been studied \cite{H9}. Without doubt, MI is the process where small perturbations are inserted into the CWs. One can also notice that the analytical investigation of the MI growth rate cannot say enough about the growing amplitude of the plane wave. As a result, numerical analysis is the most powerful tool for observing MW patterns over long periods of propagation. In what follows, we use small perturbations in the CW to establish the linearizing expression. Afterwards, we establish the MI gain, where the effects of the nonlinear terms are highlighted. To confirm our analytical investigation, we use the numerical simulation to control the exponential growth rates of the plane wave. For this, we consider the plane wave with small perturbations as having a solution of \autoref{eq:05} as:
\begin{equation}\label{eq:06}
\begin{gathered}
\psi_n=\left(F_0+F_n\right)e^{i(kn-\varpi)t},
\end{gathered}
\end{equation}
\noindent
where $F_0$ is the initial amplitude, $k$ and $\varpi$ are respectively the wave vector and angular frequency. Inserting \autoref{eq:06} into \autoref{eq:05}, gives
\begin{equation}\label{eq:07}
\begin{gathered}
i\frac {\partial }{\partial t}F_{{n}}+ \Lambda_1 F_{n+1}+ \Lambda_2F_{n-1}+ \Lambda_3F_n+\Lambda_4F^\ast_{n+1} +\Lambda_5 F^\ast_{n-1} + \Lambda_6F^\ast_{n}=0.
\end{gathered}
\end{equation}
\noindent
The parameters $\Lambda_j (j=1,...,6)$ are in the Appendix. We consider the solution of \autoref{eq:07} as follow:
\begin{equation}\label{eq:08}
\begin{gathered}
F_n=f_1\cos(Qn+\Omega{t})+if_2\sin(Qn+\Omega{t}),
\end{gathered}
\end{equation}
\noindent
where $Q$ and $\Omega$ are respectively the perturbed wave vector and angular frequency of the MI growth rate. Using \autoref{eq:08} into \autoref{eq:07}, we obtain the matrix in the form
\begin{eqnarray}\label{eq:09}
\left(\begin{array}{cccc}i\Omega-(N_{1}-N_{2}+N_{4}-N_{5})\sin(Q) &\;\; i((N_{1}+N_{2}-N_{4}-N_{5})\cos(Q)+N_{3}-N_{6})\\
\noalign{\medskip}(N_{1}+N_{2}+N_{4}+N_{5})\cos(Q)+N_{3}+N_{6}&\;\;\Omega+i(N_{1}-N_{2}-N_{4}+N_{5})\sin(Q)
\end{array}\right)\left(
\begin{array}{c} f_{1}\\ \noalign{\medskip}f_{2}\end {array}\right)= \left(
\begin{array}{c} 0\\ \noalign{\medskip}0\end{array}\right),
\end{eqnarray}
and \autoref{eq:09} can vanish only for
\begin{equation}\label{eq:10}
\begin{gathered}
\Omega^2+i(X_1-X_2)\Omega+X_1X_2+\Delta=0,
\end{gathered}
\end{equation}
\noindent
with
\begin{equation}\label{eq:11}
\begin{split}
X_1=&(N_{1}-N_{2}+N_{4}-N_{5})\sin(Q),\\
X_2=&i(N_{1}-N_{2}-N_{4}+N_{5})\sin(Q),\\
\Delta=&i((N_{1}+N_{2}-N_{4}-N_{5})\cos(Q)+N_{3}-N_{6})((N_{1}+N_{2}+N_{4}+N_{5})\cos(Q)+N_{3}+N_{6}).
\end{split}
\end{equation}
\noindent
It is worth mentioning that the MI occurs when the frequency of the modulation is complex with a non-zero imaginary part. So, the corresponding MI growth rate takes the form of
\begin{equation}\label{eq:12}
\begin{gathered}
Gain=|\Im(\Omega_{max})|.
\end{gathered}
\end{equation}
\noindent
In what follows, we highlight the impacts of the parameters of the cubic-quartic interaction potential on the MI. For this reason, the value of the cubic coupling is kept fixed along with that of $G_2.$ In \autoref{fig1}, we have depicted the variation of the MI growth rates under the effects of the quartic interaction potential. From \autoref{fig1}\,a-b, we have shown the formation of the unstable zones (bright zones) for $G_4=-0.01$. One can see that in panel (a), very slight stable zones emerge, while in panel (b), two side lobes appear with a large MI band. The higher amplitude of the plane waves is about $0.6.$ In \autoref{fig1}\,c-d, we have increase negatively the quartic parameter to  $G_4=-0.1$. In contrast to panels (a-b), the unstable zones decrease to expand the stable modes (see panel (c)), and additional bands of the MI emerge. We equally observe that the amplitude of the plane wave increases to $2.6$ and three side lobes emerge in panel (d). It results that when the quartic nonlinearity strength increases negatively, the amplitude of the perturbed plane wave and the stable modes increase together. In \autoref{fig2},\,e-f, we increased the negative nonlinear strength interaction to $G_4=-0.5$ once more. We have depicted the same scenario as in panels (c-d). The amplitude of the plane wave has increased strongly to reach $8.9$ in panel (f), and the stable modes increase. We also notice that the additional bands amplitudes increase. Once more, we have  exhibited in panels (g-h) the same behavior for $G_4=-1.$ From this analysis, it emerges that increasing negatively the quartic nonlinearity term reduces the unstable modes and increases the amplitude of the perturbed plane wave, which is a good argument for seeking numerically the evolution of the MI growth rates.\\

\noindent
Following the same procedure as in \autoref{fig1}, we have depicted unstable modes of the MI in \autoref{fig2} for $G_4=0.01,\,0.1,\,0.5,\,1.5,$ and $2.5$ in terms of $(Q,\,k).$ For $G_4=0.01$ we have shown unstable zones in panel (a), indicated by the bright zones, and three symmetric side lobes in panel (b). We increase the nonlinear term to 0.1, the unstable MI areas increase while additional bands emerge to shrink the MI bands in $k$-axis (see panel (c) and panel (d) respectively). One can observe that the amplitude of the plane wave has remained identical. To confirm the analytical predictions reported in \cite{H7}, that the quartic nonlinear term induces unstable modes, we have increase its value to $0.5$. It emerges with six peaks of the unstable modes, with small unstable lobes in the middle and a large enough amplitude of the plane wave in panels (e-f). Beside, in panel (g) and panel (h) respectively, we have depicted the same behavior, but the peak of the amplitude of the plane wave have increased strong enough to reach $200$. Without doubt, positive values of the higher nonlinear term can generate instability in a chain of atoms and could probably induce the MWs patterns during long periods of propagation.\\

\begin{figure}[H]
\begin{center}
\includegraphics[width=17cm,height=9cm]{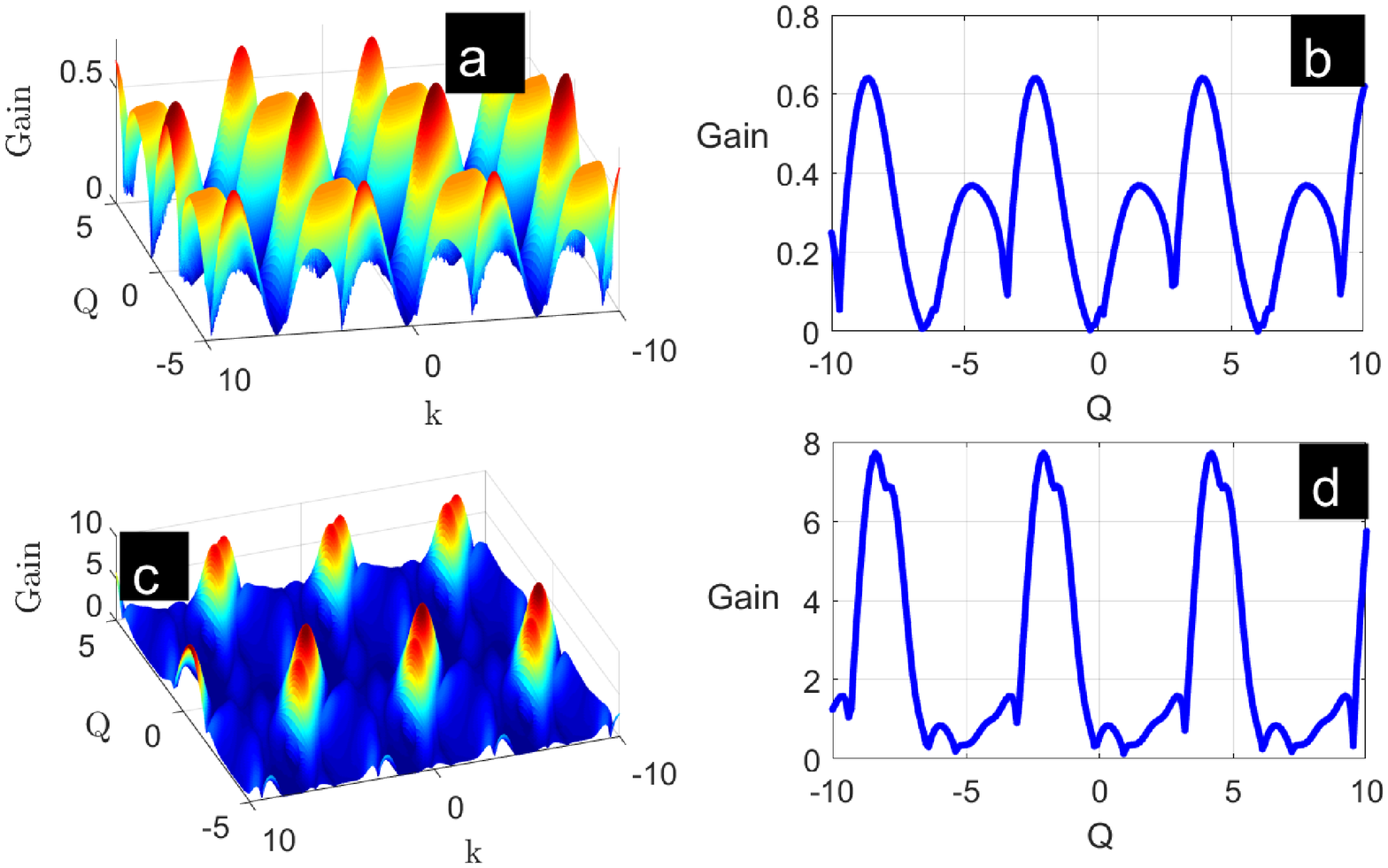}
\includegraphics[width=17cm,height=9cm]{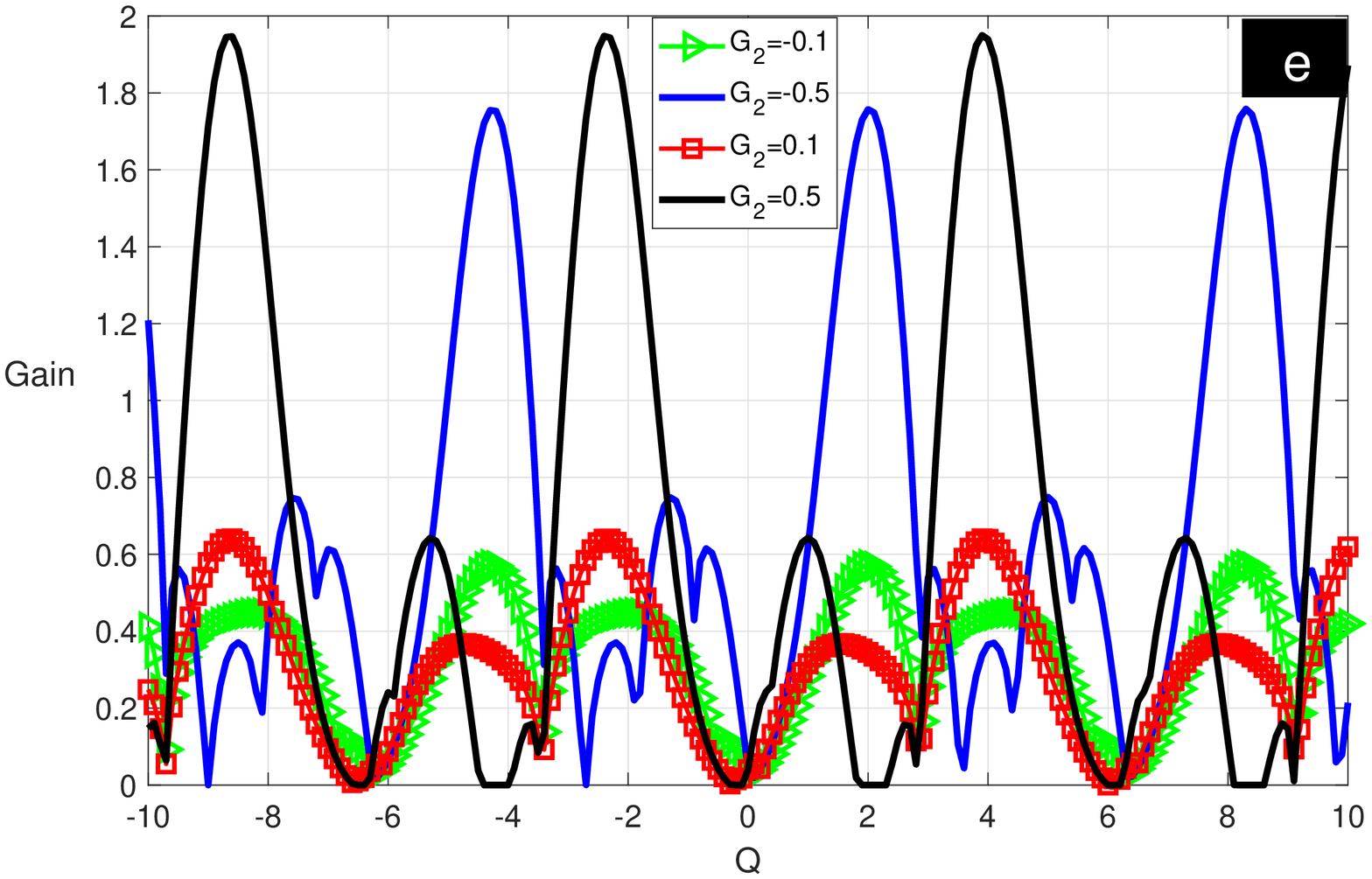}
\caption{Top panel (a-d) MI growth rate with the variation of the cubic nonlinearity interaction. (a-b) $G_3=-0.1$ and (c-d) $G_3=0.5$ Bottom panel (f) is the illustration of the MI growth rate with the variation of the dispersion term. The parameters used are respectively $G_4=0,\, G_3=0.01,\, \alpha=1.5,\, \beta=-\frac{1}{6},\, \omega_g=0.24,\, \omega=0.42,$ and $F_0=1.$}\label{fig3}
\end{center}
\end{figure}

\noindent
Since it is obvious that the quartic nonlinear term induces unstable or stable modes depending on its sign, we next turn to the effects of the cubic nonlinear strength. As a result, we consider $G_4=0$. In \autoref{fig3}\,a-d, we have depicted the variation of the unstable MI for $G_3=-0.1$ panels (a-b) and $G_3=0.5$ panels (c-d). One can observe that for negative values of the cubic term, the unstable mode is manifested by a set of two symmetrical lobes. By increasing this value to $G_3=0.5,$ the MI's stable modes increase along with the plane wave's amplitude. Another important effect of the cubic nonlinear term is observed through the unstable and stable modes in the atom's structure. Our analytical investigation has confirmed the previous predictions made by Nguetcho and co-workers. Last but not least, we used \autoref{fig3},\,e to demonstrate the manifestation of the MI growth rate with the effects of the dispersion term $G_2$. For $G_2=-0.1$ and $G_2=0.1$ respectively, the maximum amplitude of the plane wave have the same values. Meanwhile for $G_2=-0.5$ and $G_2=0.5$, one can observe that the plane wave gets large amplitude and the MI bands enlarge.
\section{Numerical investigation}\label{sec4}
In the previous section, we underlined the interplay between dispersion and high-order nonlinear terms in the structure of the one-dimensional chain of atoms. We have shown that the cubic and quartic interactions can generate unstable or stable modes as well as wave patterns. In this section, we use the numerical integration of \autoref{eq:05}.
\subsection{Modulated waves patterns}
\begin{figure}[H]
\begin{center}
\includegraphics[width=17cm,height=9cm]{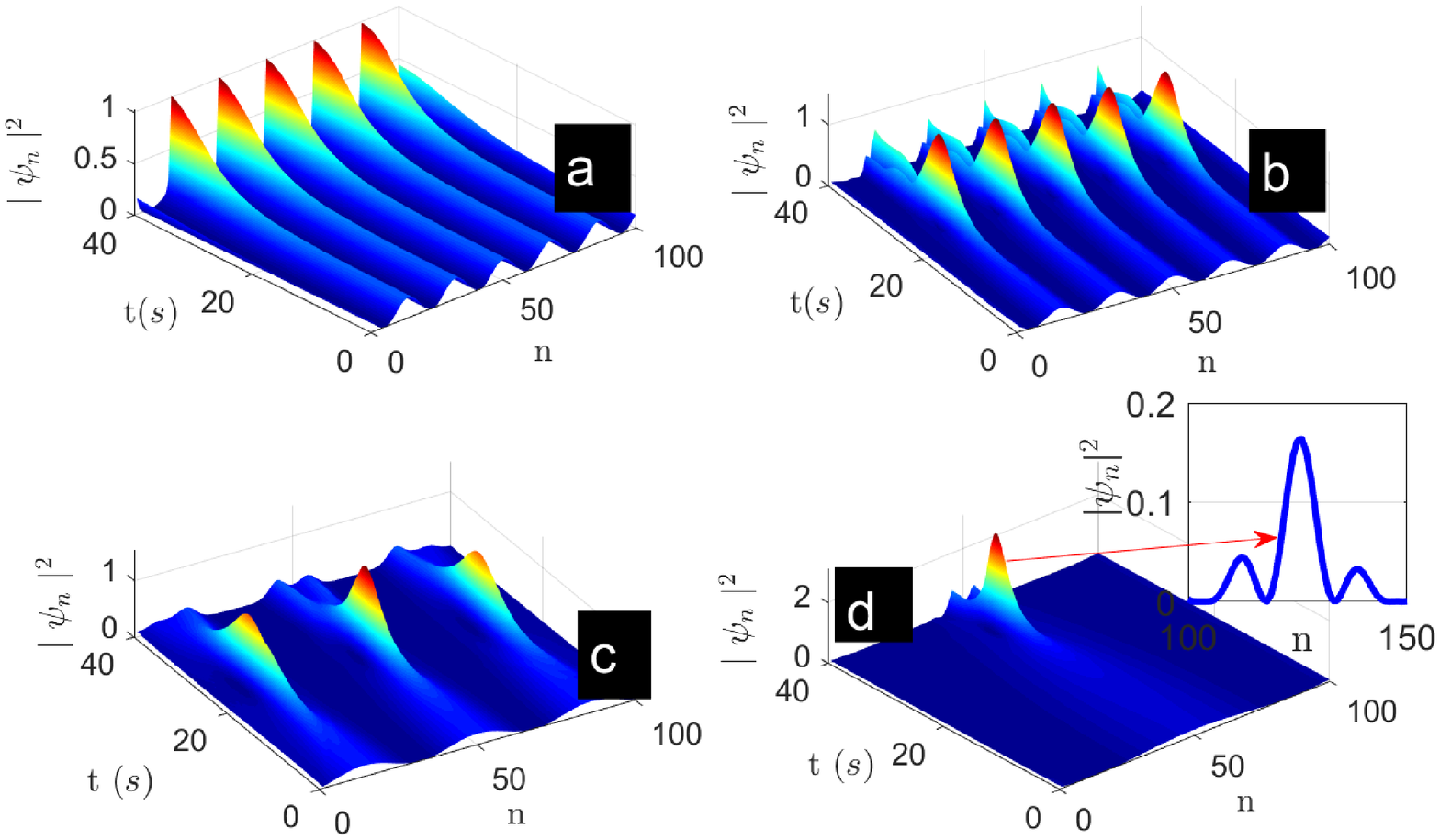}
\caption{Numerical simulation of the intensity $|\psi_n|^2$ with the variation of the quartic interaction coupling ($G_4$). (a) $G_4=0.001,\,$  (b) $G_4=0.01,\;$ (c) $G_4=0.1,\,$ and (d) $G_4=0.5.$ The parameters used are respectively $ G_2=0.01,\, G_3=0,\, \alpha=1.5,\, \beta=-\frac{1}{6},$ and $\omega_g=0.24.$ }\label{fig4}
\end{center}
\end{figure}

\begin{figure}[H]
\begin{center}
\includegraphics[width=17cm,height=9cm]{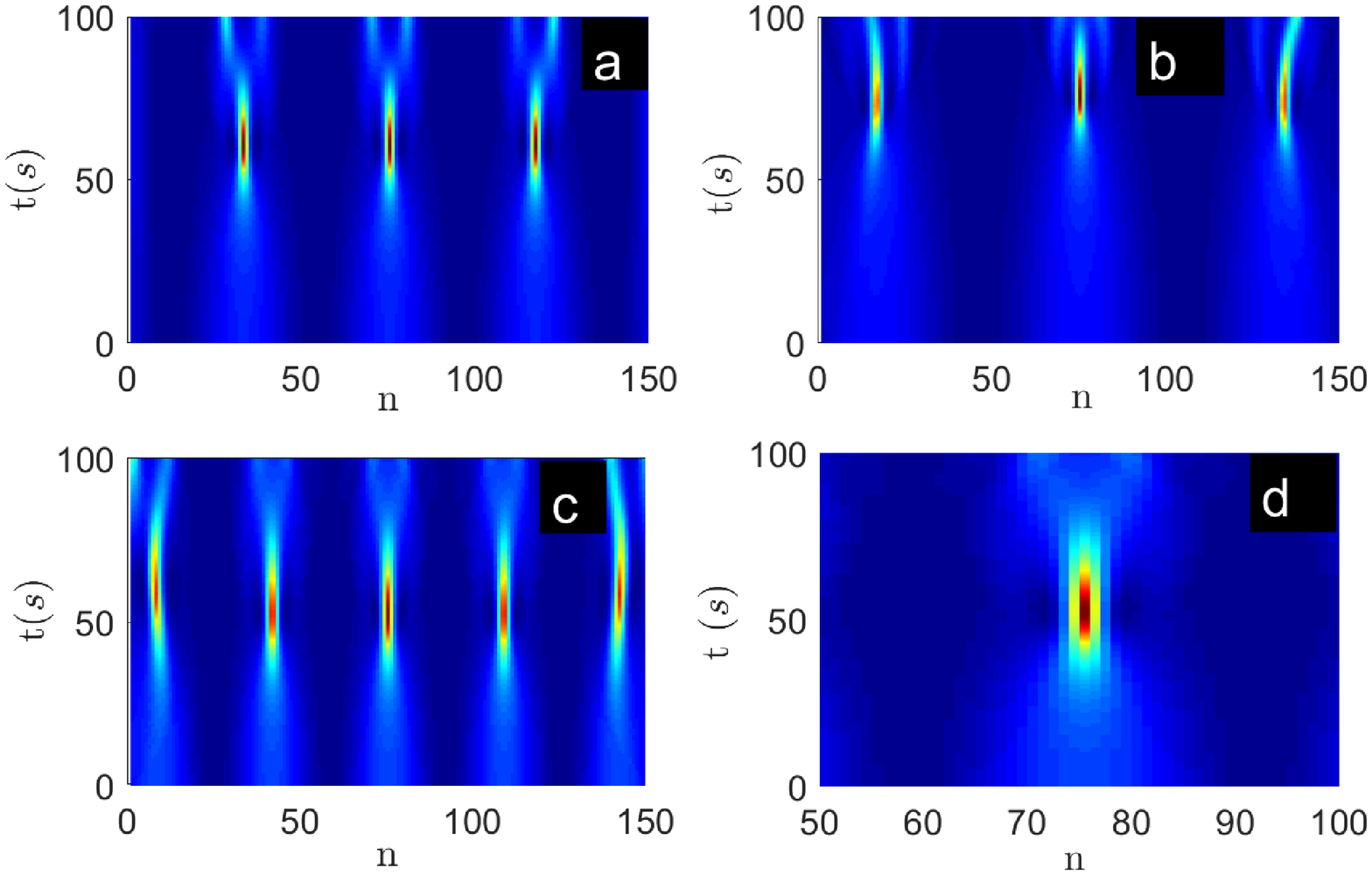}
\caption{Numerical simulation of the intensity $|\psi_n|^2$ with the variation of the quartic interaction coupling ($G_4$). (a) $G_4=-0.01,\,$  (b) $G_4=-0.1,\;$ (c) $G_4=-0.5,\,$ and (d) $G_4=-0.75.$ The parameters used are respectively $ G_2=0.01,\, G_3=0,\, \alpha=1.5,\, \beta=-\frac{1}{6},$ and $\omega_g=0.24.$ }\label{fig5}
\end{center}
\end{figure}

To say the truth, the linear stability investigation can only say so much about the long-term propagation of the CWs. To answer this preoccupation, we use the numerical integration of \autoref{eq:05}. An initial condition in the form
\begin{equation}\label{eq:13}
\begin{gathered}
\psi_n(t=0)=\psi_0\left(1+\xi\cos(Qn)\right)\cos(kn)
\end{gathered}
\end{equation}
\noindent
has been used with $\psi_0=1$ and $\xi=0.001$ to develop the MI growth rates. In what follows, attention is paid to the effects of the higher-order interaction coefficients on the development of the MI. The important aspect of this investigation is the fact that both cubic and quartic nonlinear terms of the nearest neighbor are used during the long period of the propagation of the MWs. On the other hand, the novelty of this present study lies in the inclusion of the higher-order nonlinearity that is manifested by the $G_4$ parameter. The effects of the $G_4$ nonlinear term have been revealed to be effective on the development of MI by increasing or reducing the unstable domains as well as the gain profiles \cite{H5}. Here, the effects of the higher nonlinear coupling coefficient are exhibited in \autoref{fig4}, where panel (a) shows the propagation of the trains of waves for $G_4=0.001$. By increasing the value of the higher nonlinear strength to $G_4=0.01$ in panel (b), one may observe the formation of the train of pulses with a different shape, despite the fact that the maximum amplitude of the plane wave remains constant as in panel (a). For more clarity, we have shown the same objects in panel (c) for large enough value of $G_4$ parameter. Observing closely in panel (d), one realizes a similarity with RWs, where a 2D train depicted against the background displays an Akhmediev breather despite its small amplitude. Our results seem to be in accordance with the analytical predictions where the unstable MI is developed for positive values of the nonlinear term. Otherwise, the long time propagation has open new features in the dynamics of a one-dimensional chain made of atoms, harmonically coupled to their nearest-neighbors and subjected to an external on-site potential \cite{H5}. It is equally worth to mention that in the continuum limit \autoref{eq:05} at ($k=0$) and ($k=\pi$) can turn to the nonlinear Schr\"{o}dinger equation which admits super RWs and Peregrine solitons.  Beside, in \autoref{fig5} we have shown in accordance with our analytical investigation that for negative values of the nonlinear parameter it is generated patterns waves. So, for $G_4=-0.01$,\,$-0.1$,\, $-0.5,$ and $-0.75,$ we have depicted the RWs that reinforce the argument that our structure can support Akhmediev breathers, which are related to the formation of the MI and were revealed in different studies as tools of regulation of the structures where nonlinear and dispersion terms are interplaying \cite{H5b}.\\

\begin{figure}[H]
\begin{center}
\includegraphics[width=17cm,height=9cm]{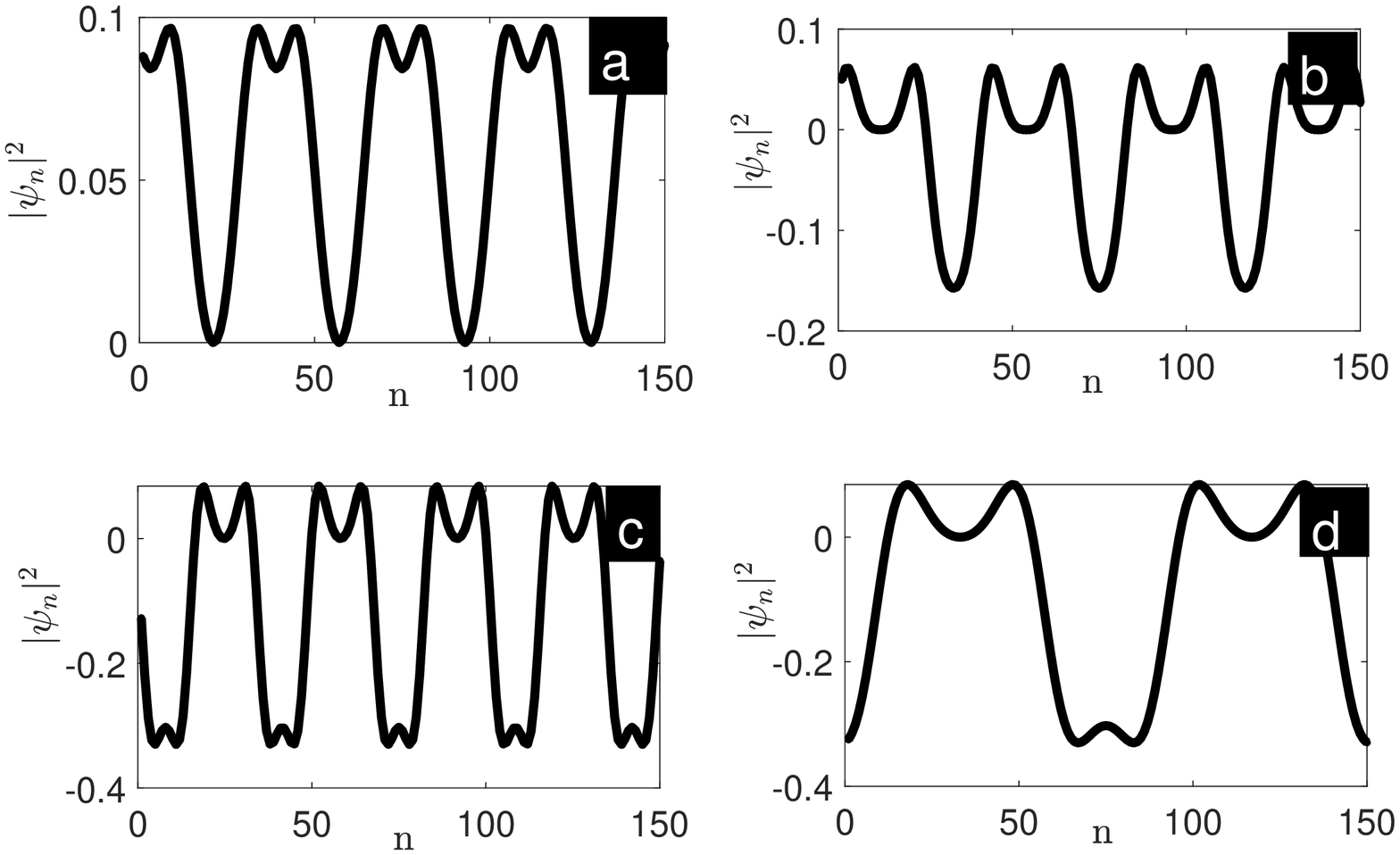}
\caption{Numerical simulation of the intensity $|\psi_n|^2$ with the variation of the cubic interaction coupling ($G_4$). (a) $G_3=-0.01,\,$  (b) $G_3=-0.5,\;$ (c) $G_3=0.01,\,$ and (d) $G_3=0.5$. The parameters used are respectively $ G_2=0.01,\, G_4=0,\, \alpha=1.5,\, \beta=-\frac{1}{6},$ and $\omega_g=0.24.$ }\label{fig6}
\end{center}
\end{figure}

\noindent
However, considering now the cubic nonlinear term gives the possibility to generate a train of pulse that comprise varied modes. In \autoref{fig6}, more precisely in panels (a,b) we have fixed $G_3=-0.01$ and $-0.5$, one observes the MWs adopt different behaviors. Following the same procedure as in panels (a,b), we point out in panels (c,d) for $G_3=0.01$ and $0.5$, the MWs patterns emerge. One results that the cubic coupling term can develop several modes when particles interaction is happened in the structure.  So, in the next section we focuss on the MW bright soliton.  Most of the studies carried out  on the effects of the higher-order interaction have used a continuum limit which is only,  our outcomes have been carried out by using a DNLEE. The obtained results have shown the robustness of this mechanism which reveals MWs with a particular properties.
\subsection{Nonlinear supratransmission}
\begin{figure}[H]
\begin{center}
\includegraphics[width=17cm,height=9cm]{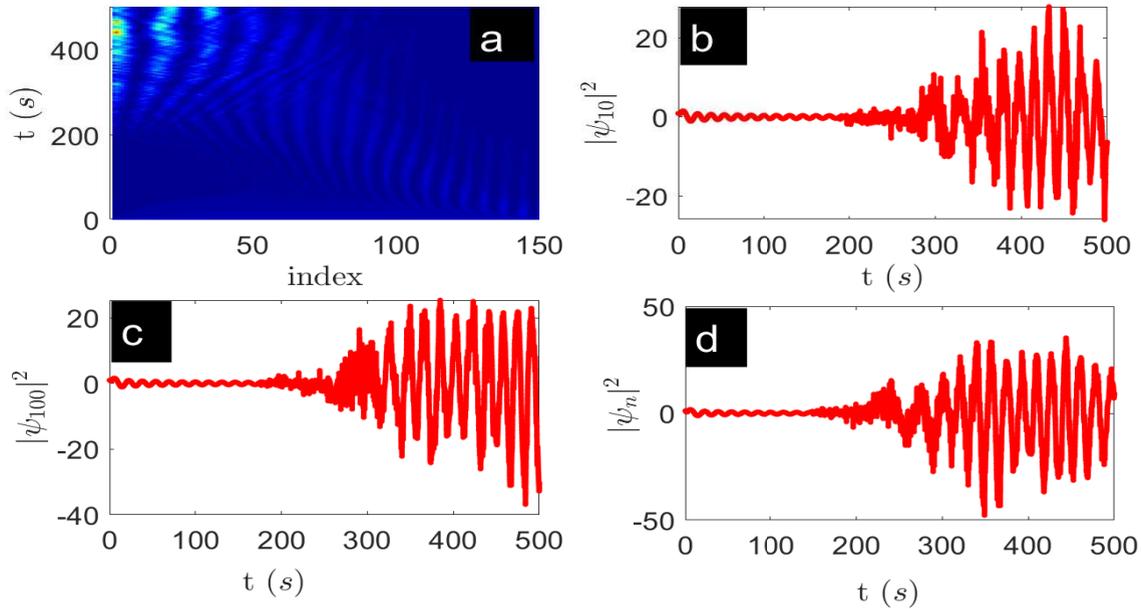}
\caption{Numerical simulation of \autoref{eq:14} submitted to the external force. (a) $A_{d_1}=1.8$  (b) $n=10,\;$ (c) $n=100,\,$ and $A_{d_1}=1.85$.  The parameters used are respectively $\omega=0.25,\, G_2=0.01,\, G_4=0.1,\, \alpha=1.5,\, \beta=-\frac{1}{6},$ and $\omega_g=0.4.$ }\label{fig7}
\end{center}
\end{figure}

In this section, we aim to submit the left and right side of \autoref{eq:05} to the external periodic force, which is different with regular mechanism of the supratransmission where only one end of the structure is driving in the FG. For this matter, we insert  $\psi_n \approx e^{i(kn-\omega t)}$ into \autoref{eq:02} and the linear dispersion frequency is $\omega=\sqrt{4G_2\sin^2(\frac{k}{2})+\omega^2_g}$. The  lower and upper frequencies are respectively $\omega_0=\omega_g$ and $\omega_{max}=\sqrt{4G_2+\omega^2_g}.$ At the center  ($k=0$) and in the limit ($k=\pi$) of the first Brillouin zone the DNLEE \autoref{eq:05} are respectively
 \begin{equation}\label{eq:14}
\begin{gathered}
-2i\omega\dot{\psi}_n+(\omega^2_g-\omega^2)\psi_n-G_2(\psi_{n+1}-2\psi_n+\psi_{n-1})+3G_3\left|\psi_{n-1}-\psi_n\right|^2\left(\psi_{n-1}-\psi\right)\\
-3G_4\left[\left|\psi_{n-1}-\psi_n\right|^2\left(\psi_{n-1}-\psi_n\right)+\left|\psi_{n+1}-\psi_n\right|^2\left(\psi_{n+1}-\psi_n\right)\right]\\
+\omega^2_g(3\beta-4\alpha^2+\frac{2\omega^2_g\alpha^2}{4\omega^2-\omega^2_g})|\psi_n|^2\psi_n=0.
\end{gathered}
\end{equation}
\noindent
and
\begin{equation}\label{eq:15}
\begin{gathered}
-2i\omega\dot{\psi}_n+(\omega^2_g-\omega^2)\psi_n+G_2(\psi_{n+1}+2\psi_n+\psi_{n-1})-3G_3\left|\psi_{n-1}+\psi_n\right|^2\left(\psi_{n-1}+\psi_n\right)\\
+3G_4\left[\left|\psi_{n-1}+\psi_n\right|^2\left(\psi_{n-1}+\psi_n\right)+\left|\psi_{n+1}+\psi_n\right|^2\left(\psi_{n+1}+\psi_n\right)\right]\\
+\omega^2_g(3\beta-4\alpha^2+\frac{2\omega^2_g\alpha^2}{4\omega^2-\omega^2_g})|\psi_n|^2\psi_n=0.
\end{gathered}
\end{equation}
\noindent
In the continuum limit the equations read at the lower FG
 \begin{equation}\label{eq:16}
\begin{gathered}
-2i\omega\dot{\psi}+(\omega^2_g-\omega^2)\psi-G_2\frac{\partial^2\psi}{\partial x^2}+\omega^2_g(3\beta-4\alpha^2+\frac{2\omega^2_g\alpha^2}{4\omega^2-\omega^2_g})|\psi|^2\psi=0.
\end{gathered}
\end{equation}
and the upper FG:
 \begin{equation}\label{eq:17}
\begin{gathered}
-2i\omega\dot{\psi}+(\omega^2_g-\omega^2+4G_2)\psi+G_2\frac{\partial^2\psi}{\partial x^2}
+\left[-24G_3+48G_4+\omega^2_g\left(3\beta-4\alpha^2+\frac{2\omega^2_g\alpha^2}{4\omega^2-\omega^2_g}\right)\right]|\psi|^2\psi=0.
\end{gathered}
\end{equation}
\noindent
The static breather solutions of Eqs.~\ref{eq:16} and \ref{eq:17} that synchronize and adjust to the driving in the end are respectively:
 \begin{equation}\label{eq:18}
\begin{gathered}
\psi_1(x,t)=A_1e^{-i(\omega-\omega_{0})t}\sech\left(\sqrt{{\frac{{\omega}^{2}+2\omega \left( \omega-\omega_{{0}}\right)-{\omega_{{g}}}^{2}}{G_{{2}}}}}(x-x_0)\right),\\
A_1=\sqrt{-\frac{8\,{\omega}^{4}+16{\omega}^{3} \left( \omega-\omega_{{0}} \right) -10{\omega}^{2}{\omega^2_{{g}}}-4\,\omega{\omega^2_{{g}}}
 \left( \omega-\omega_{{0}} \right) +2\,{\omega^4_{{g}}}}{{\omega^2_{{g}}}\left( {\frac {82{\omega}^{2}}{9}}-{\frac {19{\omega^2_{{g}}}}{6}} \right)}},
\end{gathered}
\end{equation}
and
\begin{equation}\label{eq:19}
\begin{gathered}
\psi_2(x,t)=A_2e^{-i(\omega-\omega_{max})t}\sech\left(\sqrt {{\frac {{\omega}^{2}+2\,\omega\, \left( \omega-\omega_{{max}}\right)-{\omega^2_{{g}}}-4G_{{2}}}{G_{{2}}}}}(x-x_0)\right),\\
A_2=\sqrt {-{\frac {8\,{\omega}^{4}+16\,{\omega}^{3} \left( \omega-\omega_
{{max }} \right) -10\,{\omega}^{2}{\omega^2_{{g}}}-4\,\omega\,{
\omega^2_{{g}}}\left( \omega-\omega_{{max }} \right) +2\,{\omega^4_{{g}}}-32\,{\omega}^{2}G_{{2}}+8\,G_{{2}}{\omega^2_{{g}}}}{16\,{
\omega}^{2}{\alpha}^{2}{\omega^2_{{g}}}-6\,{\alpha}^{2}{
\omega^4_{{g}}}-12\,{\omega}^{2}\beta{\omega^2_{{g}}}+3\beta_{{g}}{\omega^4_{{g}}}+96\,{\omega}^{2}G_{{3}}-192\,{\omega}^{2
}G_{{4}}-24G_{{3}}{\omega^2_{{g}}}+48G_{{4}}{\omega^2_{{g}}}}}}.
\end{gathered}
\end{equation}
\noindent
From there we derive the threshold boundary of the supratransmission in the lower and upper FGs respectively
\begin{equation}\label{eq:20}
\begin{gathered}
A_{th_1}=2\sqrt{-\frac{8\,{\omega}^{4}+16{\omega}^{3} \left( \omega-\omega_{{0}} \right) -10{\omega}^{2}{\omega^2_{{g}}}-4\,\omega{\omega^2_{{g}}}
\left( \omega-\omega_{{0}} \right) +2\,{\omega^4_{{g}}}}{{\omega^2_{{g}}}\left( {\frac {82{\omega}^{2}}{9}}-{\frac {19{\omega^2_{{g}}}}{6}} \right)}},
\end{gathered}
\end{equation}
and
\begin{equation}\label{eq:21}
\begin{gathered}
A_{th_2}=2\sqrt {-{\frac {8\,{\omega}^{4}+16\,{\omega}^{3} \left( \omega-\omega_
{{max }} \right) -10\,{\omega}^{2}{\omega^2_{{g}}}-4\,\omega\,{
\omega^2_{{g}}}\left( \omega-\omega_{{max }} \right) +2\,{\omega^4_{{g}}}-32\,{\omega}^{2}G_{{2}}+8\,G_{{2}}{\omega^2_{{g}}}}{16\,{
\omega}^{2}{\alpha}^{2}{\omega^2_{{g}}}-6\,{\alpha}^{2}{
\omega^4_{{g}}}-12\,{\omega}^{2}\beta{\omega^2_{{g}}}+3\beta_{{g}}{\omega^4_{{g}}}+96\,{\omega}^{2}G_{{3}}-192\,{\omega}^{2
}G_{{4}}-24G_{{3}}{\omega^2_{{g}}}+48G_{{4}}{\omega^2_{{g}}}}}}.
\end{gathered}
\end{equation}
\noindent
For numerical simulation, we assume the boundary condition in the form of
 \begin{equation}\label{eq:22}
\begin{gathered}
\psi_0=A_{d}\cos(\omega t),
\end{gathered}
\end{equation}
to drive \autoref{eq:14}. $A_{d}$ is the DA and $\omega$ the driven frequency (DF) with $0<\omega<\omega_0$.  In \autoref{fig7}\,a, we have shown the propagation of the local energy for the DF belonging the lower FG $\omega=0.25$ and the DA is $A_{d_1}=1.8$. For specific cells index $n=20$ and $n=100$ in  panels (b,\,c) the spatiotemporal evolution of the train of traveling wave for the left boundary is fulfilled. Increasing the DA to $1.85$, one can observe the propagation of the excited localized modes in the structure in panel (d). From the above assumption, it emerges that the model of Frenkel-Kontorova with cubic-quartic nonlinear on-site potential is opened to the nonlinear supratransmission phenomenon in the lower FG. We notice that a train of traveling wave occurs for the DA above the threshold supratransmission in the range of the propagation time $t\,\epsilon\,[200, 500].$ On the other hand, the energy goes to zero in the range of the time propagation $t\,\epsilon\,[0, 200]$ producing a transition phase  in the structure.\\

\begin{figure}[H]
\begin{center}
\includegraphics[width=17cm,height=9cm]{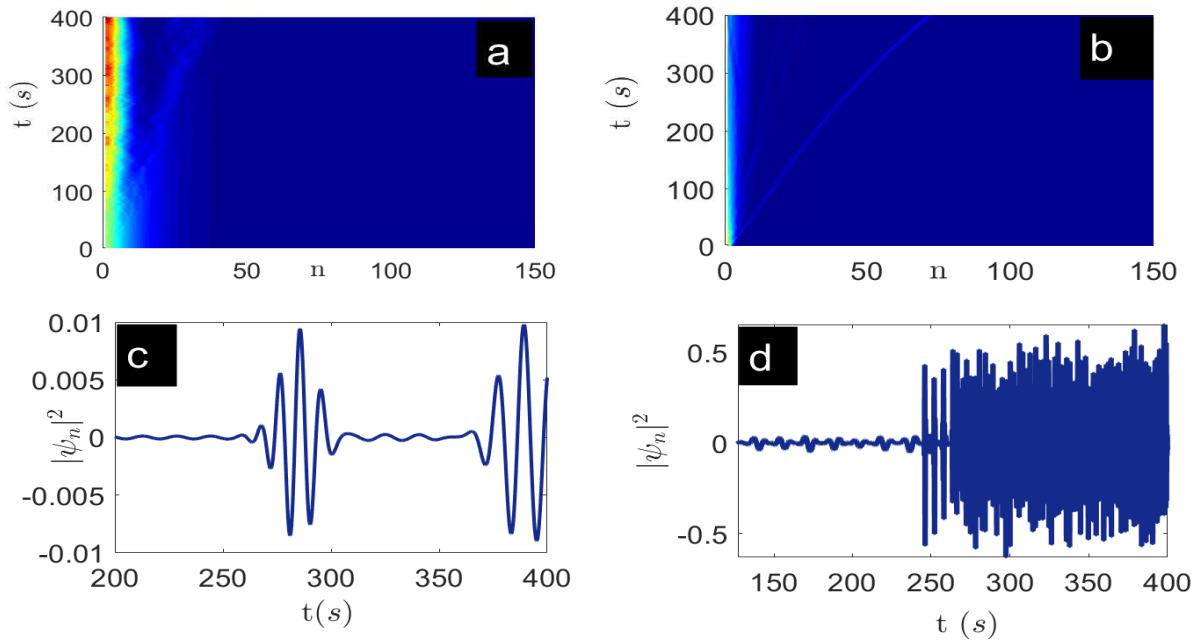}
\caption{Numerical simulation of \autoref{eq:15} submitted to the external force. (a) $A_{d_2}=0.8,$ (b) $A_{d_2}=0.85,$ (c-d) specific range of time of propagation. The parameters used are respectively $\omega_{max}=0.89,\, \omega=1,\, G_2=0.1,\, G_3=0.1,\, G_4=1,\,  \alpha=1.5,\, \beta=-\frac{1}{6},$ and $\omega_g=0.4.$ }\label{fig8}
\end{center}
\end{figure}

\noindent
Concerning the upper FG, we use the numerical integration of \autoref{eq:15}. From \autoref{fig8}\,a-b, we have depicted the propagation of the localized waves in the upper FG for the DF $\omega=1$.  For DA $A_{d_2}=0.8>A_{th_2}=0.78$, in panel (a) we show the evolution of the boundary driving of the coupled atoms with higher-order nonlinear term. In the bottom panel (c), the localized bright soliton is obtained in the range of time of propagation $t\,\epsilon\,[200,400]$. However, we have increased the DA value to 0.85, one can observe that the energy flows in upper FG  in panel (b). For specific range of propagation time, we have shown that the bright soliton turns to chaos-like motion in the structure in panel (d). This behavior corroborate our analytical prediction on the  MI growth rates where a strong value of the higher-order nonlinear term is used. On the order hand, the propagation of the localized modes occurred in the upper FG despite the fact that the DA amplitude is considered below the threshold supratransmission of the lower FG.

\section{Conclusion}\label{sec5}
In this study, we investigated the variation of the modulation instability and the behavior of the wave propagating in the forbidden bandgap. We use the one-dimensional chain of atoms harmonically coupled to their nearest neighbors. A standard multi-scale method is used to derive the discrete nonlinear evolution equation. From the linear stability, the modulation instability gain is obtained, and the impact of the cubic-quartic nonlinearity on the modulation instability leads to unstable zones as well as modulated wave patterns for certain values of the higher nonlinear term. A numerical simulation of the derived discrete nonlinear evolution equation gives birth to rogue waves and diverse types of modulated waves. We derive static breather solutions that synchronize and adjust to the drive at the center and limit of the first Brillouin zone. Thereafter, we submit one end of the discrete model to an external periodic drive. The generation of modulated waves and bright soliton is observed for driven amplitudes above the threshold of supratransmission. When the driven amplitude is increased sufficiently, the bright soliton towers into chaos-like motion in the transient range of propagation time. These results shed light on the fact that at higher orders of complexity, the modified Frenkel-Kontorova model with cubic-quartic nonlinear coupling coefficients could be used to generate rogue waves, long-lived modulated wave patterns, and chaos-like motions that are very useful for data codification.
 \section*{Appendix}
$\Lambda_1=A_{{2}}a_{{1}}b_{{1}}+2\,A_{{5}}{F_{{0}}}^{2}a_{{1}}b_{{1}} \left( a_{
{1}}b_{{1}}-1 \right)  \left( a_{{-1}}b_{{-1}}-1 \right),$\\
$\Lambda_2=A_{{2}}a_{{-1}}b_{{-1}}+2\,A_{{5}}a_{{-1}}b_{{-1}}{F_{{0}}}^{2} \left( a_{{1}}b_{{1}}-1 \right)  \left( a_{{-1}}b_{{-1}}-1 \right)\left( A_{{7}}+1 \right),$\\
$\Lambda_3= \left( -a_{{-1}}b_{{-1}}-a_{{1}}b_{{1}} \right) A_{{2}}-2\left( a_
{{1}}b_{{1}}-1 \right)  \left( a_{{-1}}b_{{-1}}-1 \right) A_{{5}}A_{{7
}}{F_{{0}}}^{2}- \left( a_{{1}}b_{{1}}-1 \right)  \left( a_{{-1}}b_{{-
1}}-1 \right)  \left( a_{{-1}}b_{{-1}}+a_{{1}}b_{{1}}+2 \right) A_{{5}
}{F_{{0}}}^{2}-A_{{7}}{F_{{0}}}^{2} \left( a_{{1}}b_{{1}}-1 \right)\left( a_{{-1}}b_{{-1}}-1 \right) ^{2}+A_{{3}}{F_{{0}}}^{2},$\\
$\Lambda_4=A_{{5}}a_{{-1}}b_{{-1}}{F_{{0}}}^{2} \left( a_{{1}}b_{{1}}-1 \right) ^{2},$\\
$\Lambda_5=A_{{5}}{F_{{0}}}^{2}a_{{1}}b_{{1}} \left( a_{{-1}}b_{{-1}}-1 \right) ^{2} \left( A_{{7}}+1 \right),$\\
$\Lambda_6=A_{{3}}{F_{{0}}}^{2}+A_{{5}} \left( A_{{7}} \left( -{F_{{0}}}^{2}{a_{{
-1}}}^{2}{b_{{-1}}}^{2}+2\,{F_{{0}}}^{2}a_{{-1}}b_{{-1}}-{F_{{0}}}^{2}
 \right) -{F_{{0}}}^{2} \left( {a_{{-1}}}^{2}{b_{{-1}}}^{2}+{a_{{1}}}^{2}{b_{{1}}}^{2}-2\,a_{{-1}}b_{{-1}}-2\,a_{{1}}b_{{1}}+2 \right)\right),$\\
$N_{{1}}=A_{{2}}a_{{1}}b_{{1}}+2\,A_{{5}}{F_{{0}}}^{2}a_{{1}}b_{{1}}\left( a_{{1}}b_{{1}}-1 \right)  \left( a_{{-1}}b_{{-1}}-1 \right),\\
N_{{2}}=A_{{2}}a_{{-1}}b_{{-1}}+2\,A_{{5}}a_{{-1}}b_{{-1}}{F_{{0}}}^{2} \left( a_{{1}}b_{{1}}-1 \right)  \left( a_{{-1}}b_{{-1}}-1 \right)\left( A_{{7}}+1 \right).\\
N_{{3}}= \left( -a_{{-1}}b_{{-1}}-a_{{1}}b_{{1}} \right) A_{{2}}-2\left( a_{{1}}b_{{1}}-1 \right)  \left( a_{{-1}}b_{{-1}}-1 \right) A_
{{5}}A_{{7}}{F_{{0}}}^{2}- \left( a_{{1}}b_{{1}}-1 \right)  \left( a_{{-1}}b_{{-1}}-1 \right)  \left( a_{{-1}}b_{{-1}}+a_{{1}}b_{{1}}+2
 \right) A_{{5}}{F_{{0}}}^{2}- \left( a_{{-1}}b_{{-1}}-1 \right) ^{2}
 \left( a_{{1}}b_{{1}}-1 \right) A_{{7}}{F_{{0}}}^{2}+A_{{3}}{F_{{0}}}^{2},\\
N_{{4}}=A_{{5}}{F_{{0}}}^{2}a_{{-1}}b_{{-1}} \left( a_{{1}}b_{{1}}-1\right) ^{2},\\
N_{{5}}=A_{{5}}{F_{{0}}}^{2}a_{{1}}b_{{1}} \left( a_{{-1}}b_{{-1}}-1\right) ^{2} \left( A_{{7}}+1 \right),\\
N_{{6}}=A_{{3}}{F_{{0}}}^{2}+A_{{5}} \left( A_{{7}} \left( -{F_{{0}}}^
{2}{a_{{-1}}}^{2}{b_{{-1}}}^{2}+2\,{F_{{0}}}^{2}a_{{-1}}b_{{-1}}-{F_{{0
}}}^{2} \right) -{F_{{0}}}^{2} \left( {a_{{-1}}}^{2}{b_{{-1}}}^{2}+{a_
{{1}}}^{2}{b_{{1}}}^{2}-2\,a_{{-1}}b_{{-1}}-2a_{{1}}b_{{1}}+2\right) \right),$\\
$A_{{1}}=\frac{\omega^{2}-\omega^2_{{g}}}{2\omega},\, A_{{2}}=\frac {G_2}{2\omega},\,
 A_{{3}}=-\frac{1}{2}\frac{{\omega}^{2} \left( 3\beta-4\,{\alpha^2_{{g}}}+{\frac {{2\omega^2_{{g}}}{\alpha}^{2}}{4{\omega}^{2}-{\omega^2_{{g}}}}} \right)}{\omega},\,A_{{5}}=\frac{3}{2}{\frac {G_{4}}{\omega}},\,
 A_{{7}}=-\frac{3}{2}\frac {G_{3}}{\omega}$,\\
$a_1=\cos(k)+i\sin(k);\,\, a_{-1}=\cos(k)+i\sin(k),$\\
$b_1=\cos(Q)+i\sin(Q);\,\, b_{-1}=\cos(Q)+i\sin(Q).$

\end{document}